\documentclass[amsmath,superscriptaddress,reprint]{revtex4-1}

\usepackage{graphicx}
\begin{document}

\title
{Ultralow mode-volume photonic crystal nanobeam cavities for high efficiency coupling to individual carbon nanotube emitters}
\author{R.~Miura}
\author{S.~Imamura}
\affiliation{Institute of Engineering Innovation, 
The University of Tokyo, Tokyo 113-8656, Japan}
\author{R.~Ohta}
\affiliation{Institute of Industrial Science, 
The University of Tokyo, Tokyo 153-8505, Japan}
\author{A.~Ishii}
\author{X.~Liu}
\author{T.~Shimada}
\affiliation{Institute of Engineering Innovation, 
The University of Tokyo, Tokyo 113-8656, Japan}
\author{S.~Iwamoto}
\author{Y.~Arakawa}
\affiliation{Institute of Industrial Science, 
The University of Tokyo, Tokyo 153-8505, Japan}
\author{Y.~K.~Kato}
\email[Corresponding author: ]{ykato@sogo.t.u-tokyo.ac.jp}
\affiliation{Institute of Engineering Innovation, 
The University of Tokyo, Tokyo 113-8656, Japan}

\begin{abstract}
We report on high efficency coupling of individual air-suspended carbon nanotubes to silicon photonic crystal nanobeam cavities. Photoluminescence images of dielectric- and air-mode cavities reflect their distinctly different mode profiles and show that fields in the air are important for coupling. We find that the air-mode cavities couple more efficiently, and estimated spontaneous emission coupling factors reach a value as high as 0.85. Our results demonstrate advantages of ultralow mode-volumes in air-mode cavities for coupling to low-dimensional nanoscale emitters.
\end{abstract}

\maketitle

Single-walled carbon nanotubes (CNTs) are known to exhibit unique optical phenomena such as multiple electron-hole pair generation \cite{Gabor:2009} and dimensionality effects on excitons \cite{Miyauchi:2013}, while their emission properties allow access to spin \cite{Stich:2014} and quantum \cite{Hofmann:2013} degrees of freedom. In order to utilize such exceptional characteristics in monolithic optical circuits, efficient coupling to photonic structures is essential. In this regard, planar cavities \cite{Xia:2008, Gaufres:2010oe, Legrand:2013, Fujiwara:2013} are not ideal as the coupling would be distributed over a continuum of modes. 

Here we demonstrate spontaneous emission coupling efficiency exceeding 85\% for a single CNT coupled to a silicon photonic crystal nanobeam cavity with an ultralow mode-volume. We take advantage of the excellent optical properties of as-grown air-suspended CNTs \cite{Mann:2007, Moritsubo:2010, Sarpkaya:2013, Barkelid:2014}, and integrate them with specially designed cavities with large fields in the air, distinctly different from the standard dielectric-mode cavities \cite{Eichenfield:2009, Gong:2010, Ohta:2011, Riedrich-Moller:2012, Hausmann:2013}. Our approach is also applicable to other low-dimensional materials, opening up a pathway for efficient use of nanoscale emitters in integrated photonics for both classical and quantum applications.

In a photonic crystal nanobeam, a periodic array of air holes is etched into a waveguide to form a photonic bandgap, which acts as a Bragg reflector. The bands below the gap are called dielectric bands because the field amplitudes are maximized within the dielectric material, while the bands above the gap are known as air bands since they have large fields in the air holes. The dielectric-band modes can be confined by locally reducing the lattice constant $a$, as the energy of the modes will become higher and the photons will be surrounded by the photonic band gap \cite{Notomi:2008, Eichenfield:2009}. Similarly, air-band modes can be confined by introducing a larger lattice constant region \cite{Zhang:2011, Quan:2011}.

We fabricate the photonic crystal nanobeam cavities from silicon-on-insulator substrates with 260~nm of top Si layer and 2~$\mu$m of buried oxide. Electron beam lithography and dry etching processes are used to form the nanobeam structure with a width of 670~nm, and the buried oxide layer is removed by wet etching. The cavities are designed to have reduced or increased lattice constant in a parabolic manner \cite{Notomi:2008,Eichenfield:2009,Ohta:2011} over 12 periods for dielectric-mode and air-mode cavities, respectively. The lattice constants and the hole sizes have been chosen to match the nanotube emission wavelengths. 

\begin{figure}[b]
\includegraphics{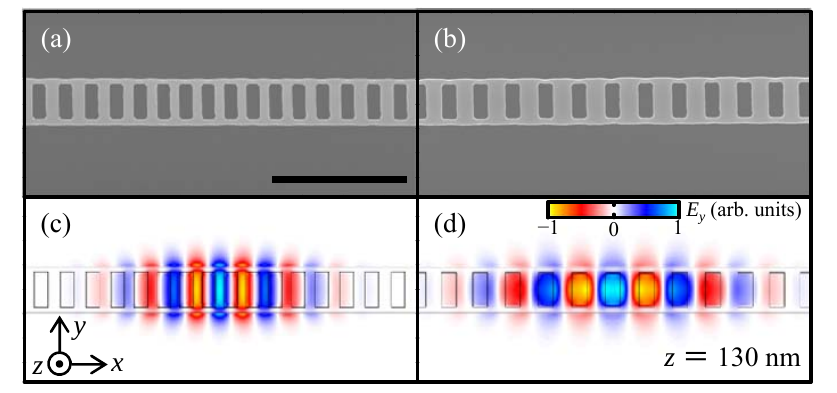}
\caption{\label{fig1}
Photonic crystal nanobeam cavities.
(a) and (b) Scanning electron micrographs of dielectric- and air-mode cavities, respectively. 
(c) and (d) Profiles of normalized $y$-component of electric fields $E_y$ at $z=130$~nm. The origin of the coordinate system is the center of the cavity. For (c), a dielectric-mode cavity with $a=390$~nm, cavity-center period of $0.84 a$, and 200~nm by 530~nm holes is used for the calculation. For (d), an air-mode cavity with $a=430$~nm, cavity-center period of $1.16 a$, and 220~nm by 510~nm holes is used.  All panels share the 2~$\mu$m scale bar in (a).
}\end{figure}

In Figs.~\ref{fig1}(a) and (b), electron microscope images of typical devices are shown. We have performed finite-difference time-domain (FDTD) calculations for these cavity structures to map out the profiles of the fundamental modes. Since nanotubes will be laying at the top surface of the nanobeam, we plot the mode profiles at that height in Figs.~\ref{fig1}(c) and (d). As expected, the dielectric-mode cavity has high field amplitudes within the Si material, with evanescent fields extending out the edges. For the air-mode cavity, the fields are mostly distributed within the air holes, with some evanescent fields as in the case of the dielectric-mode cavity. The intense fields in the air holes should be an advantage compared to cavities that confine most of the optical fields in the high-index medium \cite{Watahiki:2012, Imamura:2013}, because nanotube photoluminescence (PL) is quenched when they are in contact with the substrate \cite{Lefebvre:2003, Moritsubo:2010}. We note that both cavity modes are transverse electric modes, and therefore the polarization matches with the emission of nanotubes that are perpendicular to the nanobeams.

\begin{figure}
\includegraphics{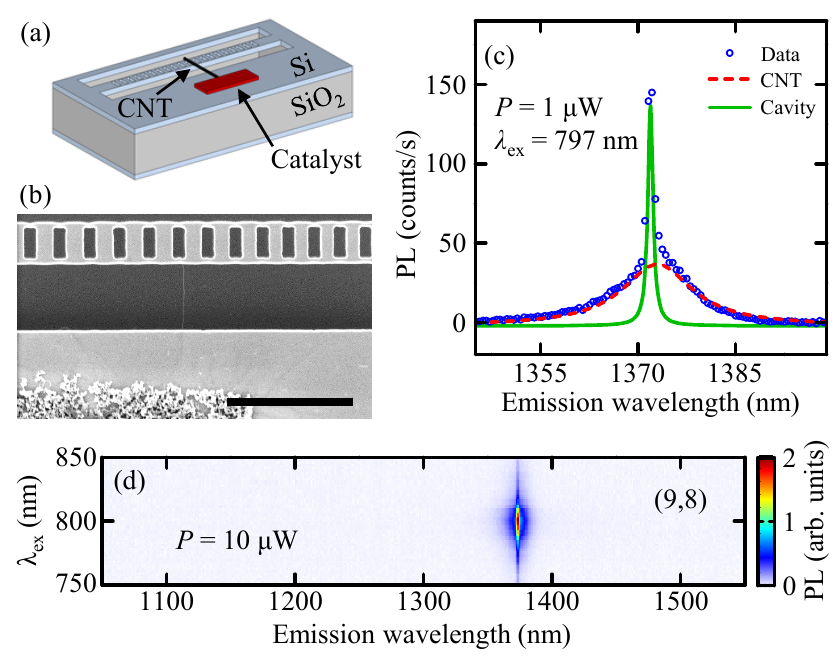}
\caption{\label{fig2}
An individual carbon nanotube coupled to a nanobeam cavity.
(a) A schematic of a device.
(b) Scanning electron microscope image of a device with a suspended nanotube. Scale bar is 2~$\mu$m.
(c) Typical PL spectrum of an air-mode device coupled to a nanotube. The dots are data and the lines are Lorentzian fits. 
(d) PL excitation map of the device shown in (c) taken with $P=10$~$\mu$W and the laser polarization perpendicular to the nanobeam.
}\end{figure}

In order to couple individual CNTs to nanobeam cavities, we have fabricated devices shown as a schematic in Fig.~\ref{fig2}(a). Catalyst particles are placed across a trench from the cavity, and we perform chemical vapor deposition to grow carbon nanotubes onto the cavities \cite{Imamura:2013}. An electron micrograph of a device after nanotube growth is shown in Fig.~\ref{fig2}(b). 

We characterize the emission properties of devices using a home-built confocal microspectroscopy system \cite{Moritsubo:2010,Watahiki:2012}. The objective lens has a numerical aperture of 0.8 and a working distance of 3.4~mm, and a pinhole corresponding to $\sim 3$~$\mu$m diameter at the image plane is used for confocal collection. The samples are excited with a wavelength-tunable continuous-wave Ti:sapphire laser, and PL is detected by an InGaAs photodiode array attached to a spectrometer. The laser polarization angle is adjusted to maximize the PL signal unless otherwise noted, and all measurements are done in air at room temperature. The samples are mounted on an automated three-dimensional stage, allowing for thousands of devices to be interrogated overnight to identify devices coupled to single nanotube emitters. 

In Fig.~\ref{fig2}(c), we present a PL spectrum from one of such devices taken with an excitation power $P=1$~$\mu$W and an excitation wavelength $\lambda_{\text{ex}} = 797$~nm. On top of the broad direct emission from the nanotube, there is a very sharp peak which is the cavity mode, indicating that the nanotube emission is optically coupled to the cavity. To further characterize the device, PL excitation spectroscopy is performed [Fig.~\ref{fig2}(d)]. We observe a single peak in the PL excitation map, demonstrating that this is an isolated single nanotube with a chirality $(9, 8)$. The intensity of the sharp cavity mode is maximized at the same wavelength as the $E_{22}$ resonance of this tube, showing that the absorption originates from the nanotube and that the cavity mode is excited by the nanotube emission.

\begin{figure}
\includegraphics{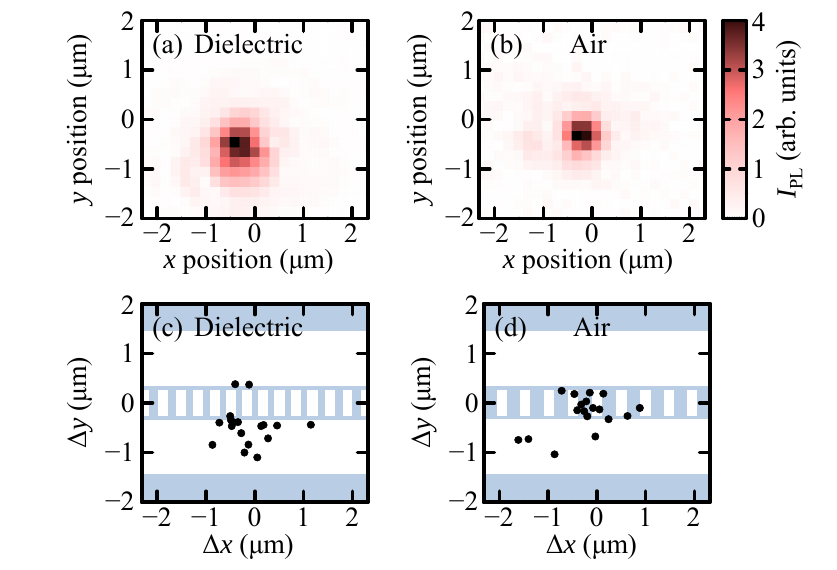}
\caption{\label{fig3}
Spatial distribution of nanotubes that show coupling.
(a) and (b) PL images of representative dielectric- and air-mode cavities, respectively, taken with   $\lambda_{\text{ex}}=800$~nm. $I_{\text{PL}}$ is obtained by integrating PL over a 0.53~nm wide spectral window centered at the cavity resonance. Excitation powers of $P=10$~$\mu$W and 20~$\mu$W are used for (a) and (b), respectively. The center of the cavities are taken as the origin of the coordinate system.
(c) and (d) Spatial distribution of PL peak intensity locations for dielectric- and air-mode cavities, respectively. The peak locations are determined by fitting to a two-dimensional Gaussian function, and they are plotted as a function of the displacement from the center of the cavity.
}\end{figure}

On those devices coupled with single CNTs, we have performed confocal PL imaging to locate the nanotube positions.  Figures~\ref{fig3}(a) and (b) show images from typical dielectric- and air-mode devices, respectively. By determining the positions of the highest PL intensity from images of more than 30 devices, the spatial displacements of the nanotubes with respect to the center of the cavities have been mapped out [Figs.~\ref{fig3}(c) and (d)]. It is possible to identify qualitative differences in the spatial distribution of nanotubes between the two types of cavities. For the dielectric-mode cavities, the nanotubes are mostly located at the edges of the nanobeam or within the trench, while for the air-mode devices, many nanotubes are on top of the nanobeam itself. 

These differences highlight the distinctness of the two types of cavities, and show that coupling occurs predominantly through fields in the air. For dielectric-mode cavities, fields within air holes in the nanobeams are weak, and evanescent fields leaking out from the edges are responsible for coupling [Fig.~\ref{fig1}(c)]. In comparison, fields in the air holes dominate for air-mode cavities [Fig.~\ref{fig1}(d)], and nanotubes are coupled when they are placed on top of the air holes.

Next, we further compare the dielectric- and air-mode cavities by analyzing the PL spectra. By performing a bi-Lorentzian fit to data [Fig.~\ref{fig2}(c)], we obtain the peak area and the linewidth for both the nanotube and the cavity emission. We let $I_{\text{cav}}$ and $I_{\text{CNT}}$ be the PL peak area of the cavity and direct CNT emission, respectively, and use $\beta^*=I_{\text{cav}}/(I_{\text{CNT}}+I_{\text{cav}})$ as a measure of the coupling efficiency. We find that the average value of $\beta^*$ for the air-mode devices is more than twice the value for the dielectric-mode cavities (Table~\ref{tab1}), consistent with the expectation from the mode profiles. We also obtain the quality factor $Q$ of the mode from the linewidth of the cavity peak, but it is likely that the observed values are limited by fabrication errors as the FDTD calculations predict $Q> 10^{5}$.

\begin{table}
\caption{\label{tab1}
$\beta^*$ and $Q$ of dielectric- and air-mode cavities measured with $P=10$~$\mu$W and  $\lambda_{\text{ex}}$ tuned to the $E_\text{22}$ resonance. The error values are standard deviations.
}
\begin{ruledtabular}
\begin{tabular}{cccc}
Cavity type & number of devices & $\beta^*$ & $Q$ \\
\hline
Dielectric & 16 & $0.18 \pm 0.16$ & $3500 \pm 1400$ \\
Air & 17 & $0.37 \pm 0.30$ & $2000 \pm 700$ \\
\end{tabular}
\end{ruledtabular}
\end{table}

\begin{figure}
\includegraphics{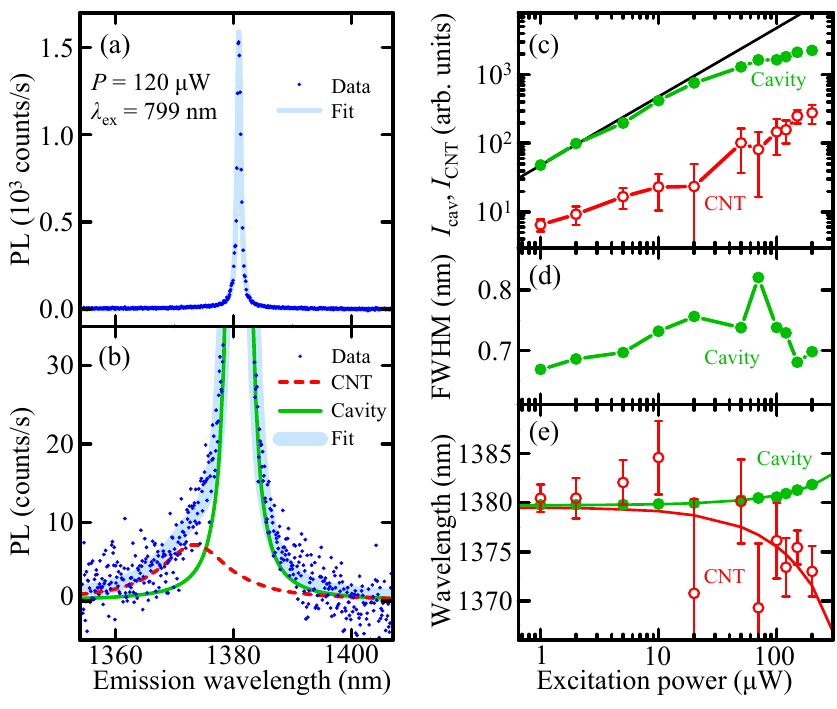}
\caption{\label{fig4}
Efficient coupling of a nanotube to an air-mode cavity.
(a) PL spectrum of an efficiently coupled device taken with  $P=120$~$\mu$W and $\lambda_{\text{ex}}=799$~nm. The dots are data and the line is a fit.
(b) An enlarged view of the low intensity region of the data shown in (a). The dots are data, thin solid line is the fit to the cavity mode, thin broken line is the fit to the CNT emission, and the thick line is the bi-Lorentzian fit.
(c) Excitation power dependence of $I_{\text{cav}}$ (filled circles)  and $I_{\text{CNT}}$ (open circles). The line is a linear function.
(d) $P$ dependence of the cavity mode full-width at half-maximum (FWHM).
(e) The center wavelength of the cavity mode (filled circles) and the nanotube peak (open circles) as a function of $P$. The solid lines are linear fits.
In (c)-(e), $\lambda_{\text{ex}}=799$~nm is used, and the error bars are smaller than the symbols for the cavity mode.
}\end{figure}

On a few air-mode cavities, we have observed very high values of $\beta^*$. In Fig.~\ref{fig4}(a), we plot the PL spectrum of a device with the highest observed $\beta^*=0.92$. It is completely dominated by the cavity mode, and the direct nanotube emission is barely observable [Fig.~\ref{fig4}(b)]. Since such a spectrum is expected for devices that are lasing, we have performed excitation power dependence measurements in search for any signs of laser oscillation.

In Fig.~\ref{fig4}(c), excitation power dependence of the PL emission intensities for the cavity and the direct CNT peaks are plotted. The cavity emission increases linearly for powers up to $\sim$20~$\mu$W and becomes slightly sublinear for higher powers, with no indication of a superlinear increase that should occur at a threshold. In addition, under lasing conditions, the direct CNT emission should saturate, because excited state population becomes constant as all the pump power is converted to cavity photon population  \cite{Yokoyama:1989}. We do not observe such saturation but the CNT peak increases linearly throughout all the powers. Furthermore, the linewidth of the cavity mode plotted in Fig.~\ref{fig4}(d) does not show the narrowing expected during lasing. From all of these observations, it is unlikely that laser oscillation is taking place. 

Under the assumption that stimulated emission is negligible, we can attribute all of the PL to spontaneous emission. Letting $\gamma_{\text{cav}}$ and $\gamma_{\text{CNT}}$ be the spontaneous emission rate into the cavity mode and all the other modes, respectively, the spontaneous emission coupling factor $\beta$ is given by
\[
\beta=\frac{\gamma_{\text{cav}}}{\gamma_{\text{CNT}}+\gamma_{\text{cav}}}
=\frac{I_{\text{cav}}}{(\eta_\text{cav}/\eta_\text{CNT})I_{\text{CNT}}+I_{\text{cav}}},
\]
where $\eta_\text{cav}$ and $\eta_\text{CNT}$ are collection efficiencies for the cavity mode and direct nanotube emission, respectively. A conservative estimate of $\beta$ is made by taking the largest possible value of the ratio $\eta_\text{cav}/\eta_\text{CNT}$. We let $\eta_\text{cav}=1$, supposing that all of the light emitted from the cavity mode into the upper hemisphere is collected by the objective. For the direct nanotube emission, we use the dipole radiation pattern as if the nanotube is emitting into free space, although we expect higher collection efficiencies because of reduced emission rate for in-plane directions caused by the photonic bandgap. This results in $\eta_\text{CNT}=0.49$ with the numerical aperture of 0.8 for the objective lens. Using these values and the fitting parameters for the data shown in Figs.~\ref{fig4}(a) and (b), we obtain $\beta=0.85$. 

It is remarkable that this value is already comparable to those for the well-established quantum-dot microcavity systems \cite{Solomon:2001, Strauf:2006, Nomura:2007, Gong:2010, Ohta:2011}, particularly because the Purcell effect is limited by the broad linewidth of nanotube emitters \cite{Exter:1996}. Nevertheless, such a high value of $\beta$ is reasonable because of ultralow mode-volume of the air-mode cavities. From the FDTD calculation  shown in Fig.~\ref{fig1}(d), mode volume $V=2.37\times10^{-2}(\lambda/n)^3$ is obtained, where $\lambda=1.38$~$\mu$m is the cavity resonance wavelength, and $n=1$ is the index of refraction for air. The maximum spontaneous emission enhancement factor is given by $F=(3\lambda^3Q_e)/(4\pi^2n^3V)=316$, where we use the quality factor of the emitter $Q_e=99$ instead of the cavity $Q$ as the nanotube emission linewidth is much wider than the cavity linewidth \cite{Exter:1996}. Even at the top surface of the nanobeam where the nanotubes are placed, an enhancement factor over 100 is obtained within the center air-hole, easily explaining the observed high $\beta$.

Finally, we would like to comment on the behavior at higher powers. In Fig.~\ref{fig4}(e), power dependence of the center wavelengths for the cavity mode and the nanotube peak are plotted. The cavity redshifts linearly with power due to heating \cite{Gong:2010}, while the nanotube peak blueshifts as observed previously \cite{Moritsubo:2010}. As a result, the cavity mode and the nanotube emission become detuned, and therefore the coupling becomes weaker. At the highest power, the detuning is 8.8~nm, and this is likely the cause of the sublinear increase of cavity mode emission at high powers [Fig.~\ref{fig4}(c)]. 

Although we did not find any signs of laser oscillations so far, with such an efficient coupling, it is expected that optimization of cavity design and fabrication would ultimately lead to lasing of individual carbon nanotubes. The air-mode cavities with ultralow mode-volumes should also allow higher coupling efficiencies for other low dimensional materials with weak dielectric screening \cite{Gan:2012, Wu:2014}.

\begin{acknowledgments}
Work supported by SCOPE, KAKENHI (24340066, 24654084, 26610080), Asahi Glass Foundation, Canon Foundation, KDDI Foundation, as well as the Project for Developing Innovation Systems and the Photon Frontier Network Program of MEXT, Japan. The devices were fabricated at the Center for Nano Lithography \& Analysis at The University of Tokyo.
\end{acknowledgments}

\bibliography{Nanobeam}

\end{document}